\documentclass[aps,twocolumn,nofootinbib]{revtex4}
\usepackage{graphicx}
\usepackage{epsfig}

\usepackage{calc}
\usepackage{ifthen}

{
{\newcommand{\gsim}{\mbox{\raisebox{-.6ex}{~$\stackrel{>}{\sim}$~}}}
{
\newcommand{\be}{\begin{equation}}
\newcommand{\ee}{\end{equation}}
\newcommand{\bea}{\begin{eqnarray}}
\newcommand{\eea}{\end{eqnarray}}

\def\GeV{{\rm \ GeV}}

\def\TeV{{\rm \ TeV}}

\begin{document}
\title{Probing Unified Origin of Dark Matter and Baryon Asymmetry at PAMELA/Fermi}
\author{Kazunori Kohri~$^1$}
\author{Anupam Mazumdar~$^{1,2}$}
\author{Narendra Sahu~$^{1}$}
\author{Philip Stephens~$^{1}$}
\affiliation{$^{1}$~Physics Department, Lancaster University, Lancaster LA1 4YB, United Kingdom\\
$^2$~Niels Bohr Institute, Copenhagen University, DK-2100, Denmark}

\begin{abstract}
We propose an unified model of dark matter and baryon asymmetry in a lepto-philic world 
above the electroweak scale. We provide an example where the inflaton decay
products subsequently generate a lepton asymmetry and a dark matter abundance with
an unique coupling in the early universe, while the present day decay of the dark matter 
through the same coupling gives rise the observed cosmic ray anomalies at PAMELA and 
Fermi Large Area Telescope.
\end{abstract}
\maketitle

The recent observed anomalies~\cite{PAMELA,cosmicray_expts,Fermi} in
the cosmic rays have given a lot of excitement. It has been shown that
there is a clean excess of absolute positron  flux in the cosmic rays
at an energy $E\gsim 50$ GeV~\cite{BMS_09}, even if the propagation
uncertainty~\cite{uncertainty} in the secondary positron flux is added
to the Galactic  background. This leaves enough motivation for
considering particle physics motivated dark matter (DM) models, such
as  annihilation~\cite{annihilation,Hamaguchi:2009jb} or
decay~\cite{decay,Hamaguchi:2009jb} of DM, as the origin of positron
excess in the cosmic rays\footnote{For astrophysical origins, see
Ref.\cite{astro-orig} and references therein.}. However the origin of the
DM, being interpreted as a long lived particle, goes beyond the standard
model (SM). Moreover, the origin of observed matter antimatter
asymmetry and the origin of inflation are also two crucial phenomena
which require the physics beyond the SM.

In particular once cosmic inflation occurs, then after inflation it
must pave the way to excite not only the observed SM quanta but also the
DM. This can be achieved minimally if the inflaton, $\phi$,  itself
carries the SM charges as in the case of the following examples which
relies on the minimal supersymmetric SM (MSSM)
setup~\cite{AEGM,AKM,ADM}. In  the MSSM case  the inflaton decays into
the SM quarks and leptons and  through thermal scatterings the
lightest supersymmetric DM particles were created which are absolutely
stable.  On the other hand if inflation does not happen in the
observable sector, for instance if it belongs to a hidden sector,  as
in the case of plethora of examples~\cite{Linde}, then the onus will
be to explain how to  generate the desired degrees of freedom, i.e. SM
baryons and dark matter abundance.

The aim of this paper is to illustrate an example where lepton asymmetry~\cite{yanagida} and DM 
abundance are generated right above the electroweak scale, i.e., ${\cal O}(100)$~GeV.
%
%
Our building block of beyond the SM physics is based on a crucial observation of cosmic ray 
anomalies~\cite{cosmicray_expts}, which may unravel the mystery of these issues in a 
{\it unifying framework} where the DM particle itself carries a net $B-L$ asymmetry and decays very
slowly to the SM particles. Note that previous attempts~\cite{KMN,darkmatter&leptogenesis_1,
darkmatter&leptogenesis_2} were made to unify dark matter and baryogenesis, but it is more 
challenging to address why the DM annihilates/decays primarily into leptons and anti-leptons 
in a {\it unifying framework}, thus explaining the observed cosmic ray anomalies at PAMELA~\cite{PAMELA} 
and Fermi~\cite{Fermi}.

For the purpose of illustration, let us augment the SM by adding a new $U(1)_{\rm B-L}$ gauge 
symmetry, and without supersymmetry. The 
anomaly free gauged $B-L$ symmetry then naturally accommodates a new fermionic dark matter
$N_L (1,0,-1)$, where the quantum numbers inside the parenthesis shows the transformation properties
of $N_L$ under the gauge group $SU(2)_L \times U(1)_Y \times U(1)_{\rm B-L}$. At a high scale the
$U(1)_{\rm B-L}$ gauge symmetry is broken by a scalar field and give mass $M_N=F v_{\rm B-L}$ 
to $N=(1/\sqrt{2})(N_L + N_L^c)$, where $``v_{\rm B-L}"$ is the vacuum expectation value (vev) of the
$U(1)_{\rm B-L}$ breaking scalar field which carries $B-L$ charges by two units and $F$ is the
coupling between $B-L$ breaking scalar field and $N_L$.

In the broken phase of the $U(1)_{\rm B-L}$ gauge symmetry the
Lagrangian involving the interactions of $N_L$, and the new massive
charged scalars, $\eta^- (1,-2,0)$ and $\chi^-(1,-2,-2)$, can be  separately
written in terms of $B-L$ conserving and $B-L$ violating parts.  The
$B-L$ violating part of the Lagrangian  is given by:
\begin{equation}
 {\cal L}_{\Delta (B-L)\neq 0}= \frac{1}{2}(M_N)_{\alpha \beta}\overline{(N_{\alpha L})^c}
  N_{\beta L} + m^2 \eta^\dagger \chi + h.c.\,,
\end{equation}
with $m^2=\mu' v_{\rm B-L}$, while the relevant $B-L$ conserving part of the Lagrangian is 
given by:
\begin{eqnarray}
{\cal L}_{\Delta(B-L)=0} &\supseteq  & M_\eta^2 \eta^\dagger \eta + M_\chi^2 \chi^\dagger \chi 
+ \mu \eta H_1 H_2 \nonumber\\
&+&h_{\alpha \beta} \eta^\dagger \overline{N_{\alpha L}} \ell_{\beta R} +
f_{\alpha \beta} \chi^\dagger \ell_{\alpha L} \ell_{\beta L}+ h.c.\,
\label{Lagrangian}
\end{eqnarray}
where the indices $\alpha, \beta= e, \mu, \tau$ represent the flavor basis of the SM fermion
fields. In equation (\ref{Lagrangian}), $\ell_L (2,-1,-1)$ and $\ell_R(1,-2,-1)$ represent the 
lepton doublet and singlet respectively, while $H_1 (2,1,0)$, $H_2 (2,1,0)$ are two doublet Higgses 
which couple to up and down sector of SM fermions.

An important point to note is that the mass term, $m^2 \eta^\dagger \chi$, which violates $B-L$ by 
two units, gives rise a mixing between $\eta$ and $\chi$. Due to this mixing the Majorana fermion 
$N$ will be an {\it unstable leptonic} DM. Therefore, the decay products of $N$ are only 
SM leptons/antileptons. If the lifetime of $N$ is about ${\cal O} (10^{25})$ s or so, 
then the decay products of $N$ can naturally account for the observed $e^\pm$ excesses 
at PAMELA and Fermi. This observation will place non-trivial constraint, such as $m\ll M_\eta, 
M_\chi$. Further we note that the only coupling $``h"$ is responsible for the production of a 
net lepton asymmetry and DM in the early universe, while the three body decay of DM 
through the same coupling at current epoch gives rise to the observed anomalies at PAMELA 
and Fermi.

\noindent
{\it \underline{Baryon asymmetry}}:\\
Let us now consider how can we explain the observed matter-anti-matter asymmetry in our setup.
A natural possibility is to generate a lepton asymmetry~\cite{darkmatter&leptogenesis_2,hambye&ma} 
from the out-of-equilibrium decay of $\eta^-$ field. Similar to Dirac leptogenesis~\cite{dirac_lep}, 
here we will argue that the required baryon asymmetry can be generated from a conserved $B-L$ number. 
In order to generate the baryon asymmetry we need following three steps:

\begin{enumerate}

\item{At first the CP-violating out-of-equilibrium decay of $\eta^-$
must generate an equal and opposite $B-L$ asymmetry between $N_L$ and $\ell_R$. These 
two asymmetries should not equilibrate above the electro weak (EW) phase transition.}

\item{Above the EW-phase transition the $B-L$ asymmetry stored in $\ell_R$ gets 
transferred to $\ell_L$, while keeping an equal and opposite $B-L$ asymmetry in $N_L$.}

\item{The $B-L$ asymmetry stored in $\ell_L$ then gets converted to a net baryon
asymmetry in the presence of $SU(2)_L$ sphalerons, while keeping the $B-L$ asymmetry stored 
in $N_L$ intact.}

\end{enumerate}
Note that all three steps happen right above the EW scale. Since $\eta^-$ is neutral 
under $B-L$, it can decay to a pair of lepton ($\ell_R$) and antilepton ($\overline{N_L}$). 
Note that the decay of $\eta^-$ can not produce any lepton asymmetry since its decay does 
not violate any lepton number. However, if there are at least two $\eta^-$ fields, say 
$\eta_1^\pm$ and $\eta_2^\pm$ then there can be CP violation in the decay of $\eta^-$ fields. 
In their mass basis, spanned by $\psi_1^\pm$ and $\psi_2^\pm$, the lightest $\psi$ field, 
say $\psi_1^\pm$, can generate a net CP asymmetry through the interference of tree level and 
self energy correction diagram~\cite{ma&sarkar}. The CP asymmetry is then given by
\begin{equation}
\epsilon_1=\frac{ {\mathrm Im}\left[ (\mu_1 \mu_2^*) \sum_{ij} h^1_{ij}
h^{*2}_{ij} \right]}{16 \pi^2 (M_{\eta_2}^2-M_{\eta_1}^2)} \left[ \frac{M_{\eta_1}}
{\Gamma_{\eta_1}}\right]\,,
\label{cpasymmetry}
\end{equation}
where 
\begin{equation}
\Gamma_{\eta_1}=\frac{1}{8 \pi  M_{\eta_2}}\left( \mu_1 \mu_2^* + M_{\eta_1} M_{\eta_2} 
\sum_{i,j} h^1_{ij} h^{2*}_{ij} \right)\,.
\label{eta_decayrate}
\end{equation}
Now assuming 
\begin{eqnarray}
& \frac{ M_{\eta_1} M_{\eta_2}} {(M_{\eta_2}^2-M_{\eta_1}^2)} & ={\cal O}(1),~~~
\frac{\mu_1 \mu_2}{ M_{\eta_1}M_{\eta_2}}= {\cal O}(1) \nonumber\\
&&{\rm and}~~~~h^1_{ij} \simeq h^2_{ij}={\cal O}(10^{-2})\,, 
\label{assumption}
\end{eqnarray} 
we get from Eqns. (\ref{cpasymmetry}), (\ref{eta_decayrate}) and (\ref{assumption}) 
the CP asymmetry $\epsilon_1 \simeq 10^{-5}$. Due to the CP violation the decay of 
$\psi_1^\pm$ generates an equal and opposite $B-L$ asymmetry between $N_L$ and $\ell_R$. 
Since the interaction between $N_L$ and $\ell_R$ through the coupling `h' is already 
gone out-of-equilibrium, the asymmetries between them don't equilibrate any more at the 
required scale of ${\cal O} (100)$ GeV. On the other hand, the lepton number conserving process: 
$\ell_R \ell_R^c \leftrightarrow \ell_L \ell_L^c$, mediated via the SM Higgs, remains 
in thermal equilibrium above the electroweak phase transition. As a result the $B-L$ asymmetry 
stored in $\ell_R$ gets transferred to $\ell_L$ through this L-number conserving process, 
while leaving an equal and opposite $B-L$ asymmetry in $N_L$. The transportation of 
$B-L$ asymmetry from $\ell_R$ to $\ell_L$ can be understood as follows. Let us define 
the chemical potential associated with the $\ell_R$ field as $\mu_{eR}=\mu_0 + \mu_{\rm BL}$, 
where $\mu_{\rm BL}$ is the chemical potential contributing to $B-L$ asymmetry and $\mu_0$ 
is independent of $B-L$. Hence at equilibrium we have the chemical potential associated 
with $\ell_L$ is given by $\mu_{eL}=\mu_{eR} + \mu_{H}=\mu_{\rm BL} + \mu_0 + \mu_{H}$. 
Thus we see that the same chemical potential is associated with $\ell_L$ as of $\ell_R$. 
Therefore, the net $B-L$ asymmetry stored in $\ell_R$ can be passed on to $\ell_L$. Since the
$SU(2)_L$ sphalerons are in thermal equilibrium at a scale above 100 GeV, the $B-L$
asymmetry stored in $\ell_L$ can be converted to a net baryon asymmetry, while an
equal and opposite $B-L$ asymmetry will remain in $N_L$. The two asymmetries will equilibrate 
when $N_L$ will decay through the $B-L$ violating process. The net $B-L$ asymmetry thus produced 
can be given as:
\begin{equation}
\eta_{B-L}=\frac{3}{4}B_\eta \epsilon_1 \frac{T_R}{m_\phi}\,,
\end{equation}
where $B_\eta$ is the inflaton branching ratio, which is of order, ${\cal O}(1)$, and 
$T_{R}\gsim 100$~GeV is the reheat temperature of the universe and $m_{\phi}$ is the inflaton mass.

The conversion of lepton asymmetry to the baryon 
asymmetry is obtained by $\eta_B=(28/79)\eta_{B-L}$. For $T_R/m_\phi\approx 10^{-4}$ 
and $\epsilon_1\approx 10^{-5}$, we can achieve the observed baryon asymmetry $\eta_B\approx 
{\cal O}(10^{-10})$. A crucial point to note here is that the lepton asymmetry is virtually 
created by a non-thermal decay of the inflaton decay products, $\eta$ and $\chi$, which we will 
discuss below.

An obvious danger of washing out this asymmetry comes from the $B-L$ violating process
$N_L \ell_R \rightarrow \ell_L \ell_{L}$ through the mixing between $\eta$ and $\chi$.
However, this process is suppressed by a factor $(m^2/M_\eta^2 M_\chi^2)^2$ for $m\ll M_\eta, 
M_\chi$, and hence it cannot compete with the Hubble expansion parameter at $T_R \sim 
100 $ GeV. Hence the net $B-L$ asymmetry produced by the decay of $\eta$ will be converted 
to the required baryon asymmetry without suffering any washout.\\

\noindent
{\it \underline{Dark matter abundance}:}\\
Next we discuss the number density of the lightest $N_L$ to check if it satisfies the observed 
DM abundance. It turns out that a large abundance of $N_L$ will be produced non-thermally by 
the decay of $\eta$, which is also non-thermally produced by the inflaton decay. 

As we will argue below, the annihilation cross section of $N_L$ is larger than the canonical one, $\langle
\sigma|v| \rangle_{{\rm c}} \sim 3\times 10^{-26} {\rm cm}^{3} {\rm
s}^{-1}$. In this case the final abundance of the thermal component is much
smaller than the observational value,
\begin{equation}
Y_{\rm DM} \equiv \frac{n_{\rm DM}}{s} 
= 4 \times 10^{-13} \left(\frac{1 \TeV}{M_{\rm DM}} \right) \left(
\frac{\Omega_{\rm DM} h^2}{0.11}\right),
\label{eq:obsDM}
\end{equation}
where $M_{\rm DM}$ is the DM mass, $\Omega_{\rm DM}$ is the density parameter of 
the DM with $h$ being the normalized Hubble constant.

Since we consider a case when the inflaton, $\phi$, decays well after the standard
freeze-out epoch of the thermally-produced $N_L$, i.e., at cosmic temperature $T 
\lesssim M_{N}/25$. Then the yield value of $N_{L}$ is estimated by
\begin{equation}
Y_{N_{L}} \equiv \frac{n_{N_L}}{s} \simeq \frac{3}{4} B_\eta \frac{T_R}{m_\phi}\,.
\end{equation}
For $T_R/m_\phi \sim 10^{-4}$, as required by the lepton asymmetry, we find that the relic 
abundance of $N_L$ is given by $Y_{N_{L}} \approx 10^{-5}$, which is much
larger than  the observed DM abundance (\ref{eq:obsDM}).  

\begin{figure}[ht]
\begin{center}
\epsfig{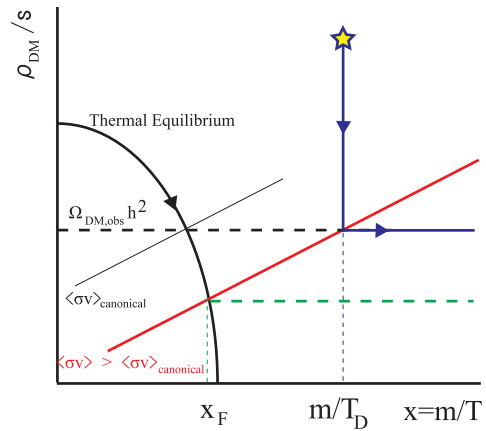}
\caption{We illustrate the time evolution of $\rho_{\rm DM}/s$ as a function of $x
\equiv M_{\rm DM}/T$, where $\rho_{\rm DM}/s \equiv M_{\rm DM} Y_{\rm
DM}$. $x_{\rm F}$ is the thermal-freezeout value of $x$ ($\sim$
25). $T_{\rm D}$  is the cosmic temperature when DM is non-thermally
produced ($T_{\rm D}=T_{R}$ in the current model). The curved line traces the
thermal-equilibrium value of DM when the thermal-production process is
important, i.e., $x < x_{F}$, which was omitted in
Eq.~(\ref{eq:boltzmann}). The diagonal thick (thin) solid line
represents freeze-out value when $\langle \sigma|v|\rangle$ is larger
than (or equal to) the canonical value. The upper-horizontal dashed line
shows the observational abundance. In the current model, the thermal
component (lower-horizontal dashed line) is smaller than the
observational value. The star symbol represents the point when the
initial value was put at $T=T_{\rm D}=T_{R}$.  Just after this non-thermal
production, the DM annihilates immediately, and its abundance is  reduced
to the crossing point between the diagonal thick-solid line and the
vertical line, from then it becomes constant.}
\label{fig:freezeout}
\end{center}
\end{figure}

However thanks to the gauge coupling of $N_L$, such that $N_L$ can now annihilate into the SM
fermions through the exchange of $Z_{\rm B-L}$ gauge boson after its
non-thermal production. We can then obtain the final abundance of $N_L$ by
solving the Boltzmann equations:
\begin{eqnarray}
\frac{dn_\eta}{dt}+3 n_\eta H=-\Gamma_\eta n_\eta \,,\nonumber\\
\frac{dn_{N_{L}}}{dt} + 3 n_{N_{L}} H = -\langle \sigma|v| \rangle
n_{N_{L}}^2 + \Gamma_\eta n_\eta\,,
\label{eq:boltzmann}
\end{eqnarray}
where $\langle \sigma|v|\rangle\approx  (1/4\pi)M_N^2/v_{\rm B-L}^4$,
and we have omitted the production term from the thermal bath,
$+\langle \sigma|v| \rangle n_{N_L,{\rm eq}}^{2}$ in the right-hand side of
the second line. Then we approximately obtain,
\begin{equation}
Y_{N_{L}}  \simeq  \frac{3 H}{\langle \sigma|v|\rangle  s}.
\label{eq:Yfreezeout}
\end{equation}
The right-hand side of Eq.~(\ref{eq:Yfreezeout}) is 
$\propto 1/T_{R}$, which  means that the late-time decay induces a larger 
freeze-out value. In Fig.~\ref{fig:freezeout} we illustrate
such a non-thermal production and/or further annihilation mechanism.

By equating $Y_{N_{L}} \simeq Y_{\rm DM}$  we get a constraint on the
$B-L$ breaking scale with satisfying the observational DM density to be
\begin{eqnarray}
v_{\rm B-L} \simeq 5 \times 10^3 \GeV
&&\left( \frac{\Omega_{\rm DM} h^2 }{0.11} \right)^{1/4}
 \left( \frac{M_N}{3 \TeV} \right)^{1/4}\, \nonumber \\
&&\times \left(\frac{T_R}{100 \GeV} \right)^{1/4}.
\end{eqnarray}
This gives $\langle \sigma|v|\rangle \sim {\cal O}(10^{-25}) - {\cal
O}(10^{-24})~{\rm cm}^{3}{\rm s}^{-1}$ for $M_{N} \sim O({\rm TeV})$,
which is larger than the canonical annihilation cross section and
makes the thermal component sub dominant. In turn, we understand this
mechanism intuitively by using Fig.~\ref{fig:freezeout} as follows. If
we specify a $U(1)_{B-L}$ breaking scale ($\sim $ TeV in this model) or an
annihilation cross section, the model necessarily has a crossing point
between the diagonal and the horizontal lines at $x = M_{\rm
DM}/T_{R}$ shown in Fig.~\ref{fig:freezeout}, which gives the right
observational abundance of DM and its production epoch. In the current
model, we demand the cross section to be larger than the canonical
value and $T_{R}$ to be larger than 100 GeV.

Note that the annihilation process $\bar{N_L} N_L \rightarrow \bar{f} f$ is 
a $B-L$ number conserving process and therefore does not transfer any $B-L$ asymmetry 
to the SM fermions. As a result the $B-L$ asymmetry produced via the decay
of $\eta^-$ will survive until far below the electro-weak phase transition.

\begin{figure}[htbp]
\begin{center}
\epsfig{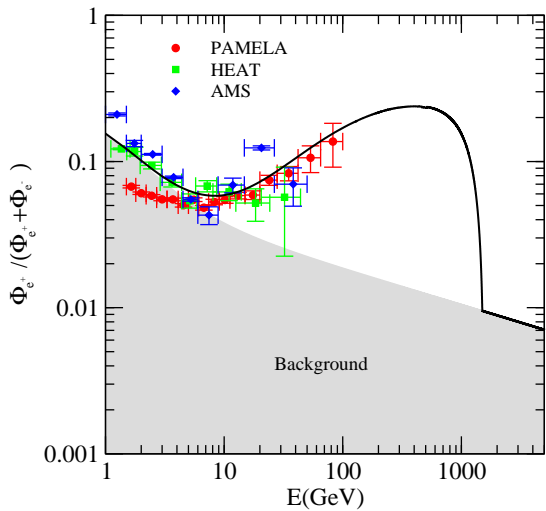}
\caption{Positron excess from $N_L \rightarrow \tau^- \tau^+ \bar{\nu} $ with $M_N=3$ TeV. The 
fragmentation function has been calculated using PYTHIA~\cite{Sjostrand:2006za}.}
\label{fig-2}
\end{center}
\end{figure}

\begin{figure}[htbp]
\begin{center}
\epsfig{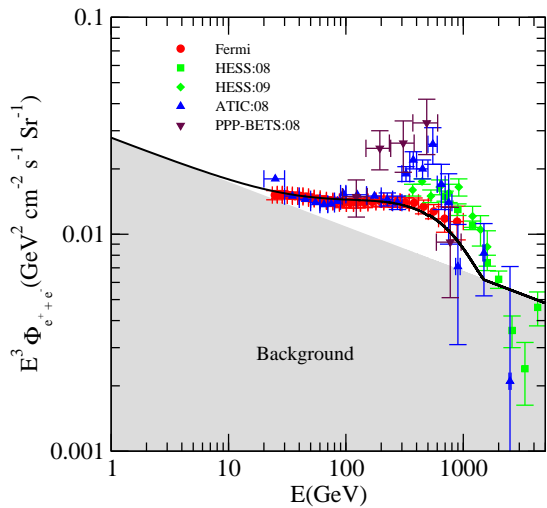}
\caption{$e^\pm $ excess from $ N_L \rightarrow \tau^- \tau^+ \bar{\nu} $ with $M_N=3$ TeV. The 
fragmentation function has been calculated using PYTHIA~\cite{Sjostrand:2006za}.}
\label{fig-3}
\end{center}
\end{figure}

Since $B-L$ is already broken, the lightest $N_L$ is no more stable. It will decay through
the three body process: $N_L \rightarrow e_{\alpha R}^- e_{\beta L}^+ \overline{\nu}_{\gamma L}$, with 
$\beta \neq \gamma$, through the mixing of $\eta$ and $\chi$. Since the coupling of $\chi$ to 
two lepton doublets is antisymmetric, i.e., $\beta \neq \gamma$, the decay of $N_L$ is not 
necessarily to be flavour conserving. In particular the decay mode: $N_L \rightarrow \tau_{R}^- 
\tau_{L}^+ \overline{\nu}_{e L} (\overline{\nu}_{\mu L})$, violates $L_e$ ($L_\mu$) by one unit 
while it violates $L=L_e + L_\mu + L_\tau$ by two units. In the mass basis of $N_L$ the life time 
can be estimated to be 
\begin{eqnarray}
\tau_{N}  &=& 2.0 \times 10^{25} {\rm s} \left( \frac{10^{-2}}{h}\right)^2 \left( \frac{10^{-7}}{f} 
\right)^2 \nonumber\\
&& \left(\frac{50 \GeV}{m} \right)^4 \left( \frac{m_\phi}{10^6 \GeV} \right)^8 
\left( \frac{1 \TeV}{M_{N_L}}\right)^5\,,
\end{eqnarray}
where we assume that $M_\eta \simeq M_\chi \approx m_\phi$ in order to get a lower limit 
on the lifetime of $N_L$. The prolonged life time of $N_L$ may explain the current cosmic 
ray anomalies observed by PAMELA~\cite{PAMELA} and Fermi~\cite{Fermi}. The electron and positron 
energy spectrum can be estimated by using the same set-up as in Ref.~\cite{decay}. In Figs. \ref{fig-2} 
and \ref{fig-3} we have shown the integrated $e^\pm$ fluxes in a typical decay mode: $N_L \rightarrow 
\tau^- \tau^+ \bar{\nu} $ up to the maximum available energy $M_N/2$ for $\tau_N=4.0\times 10^{25}$ 
secs. From there it can be seen that the decay of $N_L$ can nicely explain the observed positron excess 
at PAMELA and $e^\pm$ excesses at Fermi. While doing so we assume that the branching fraction in the 
decay of $N-L$ to $\tau^- \tau^+ \bar{\nu}$ is significantly larger than the other viable decay modes: 
$N_L \rightarrow \mu^- \mu^+ \bar{\nu} $ and $N_L \rightarrow e^- e^+ \bar{\nu} $. However, if the decay 
rate of $N_L \rightarrow \mu^- \mu^+ \bar{\nu} $ is comparable to  $N_L \rightarrow \tau^- \tau^+ \bar{\nu} $ 
then it can explain the observed anomalies at PAMELA and Fermi, while the decay mode: 
$N_L \rightarrow e^- e^+ \bar{\nu} $ produces larger $e^+ + e^-$ fluxes at Fermi and therefore 
unfavorable.      

Another potential signature of this scenario is the emission of energetic neutrinos from 
the Galactic center~\cite{GCnu} which can be checked by future experiments such as IceCube 
DeepCore~\cite{Cowen:2008zz} and KM3NeT~\cite{Kappes:2007ci}. We will come back to this 
issue in a future work~\cite{kmsp}. \\


\noindent
{\it \underline{Inflation and reheating}}:\\
So far we have not discussed anything about inflation sector. Let us now consider a hidden sector 
inflaton $\phi(1,0,0)$ which excites the observed DM abundance and SM leptons during the process of reheating 
which occurs after the end of inflation. Let us consider a simple toy model of inflation, where 
the inflaton potential admits a point of inflection, see ~\cite{AEGM,AKM,KMN}:
\begin{equation}
V(\phi)\sim \frac{m_{\phi}^2}{2}\phi^2-\frac{A\kappa}{6\sqrt{3}}\phi^3+\frac{\kappa^2}{12}\phi^4\,,
\end{equation}
where $A\approx 4m_{\phi}$ and $\kappa\sim 10^{-10}$. Inflation can happen
near a point of inflection given by $\phi_0\sim \sqrt{3}m_{\phi}/\kappa\sim 10^{16}$~GeV
with an Hubble expansion rate, $H_{inf}\sim (m_{\phi}^2/\kappa M_P)\sim
10^{4}$~GeV.  The amplitude of the density perturbations will be give
by:  $\delta_H \approx (1/5\pi)(H_{inf}^2/\dot\phi)\sim (\kappa^2M_P/3m_{\phi}){\cal N}^2\sim 
10^{-5}$, where the  number of e-foldings is given by: ${\cal N}^2\sim 10^{3}$~\cite{Burgess}. 
One of the dynamical properties of an inflection point inflation is that the spectral tilt
can be matched in a desired observable range: $0.92< n_s< 1.0$ for the above 
parameters~\cite{AEGM,AKM,KMN,LYTH}.

The inflaton decays into heavy charged scalars $\eta$ and $\chi$. Let $B_\eta$ be the 
branching fraction in the decay of $\phi$ to $\eta^\pm$ and $B_\chi$ be the branching 
fraction in the decay of $\phi$ to $\chi^\pm$. As we discussed above the charged scalars 
$\eta^\pm$ and $\chi^\pm$ couple to the SM degrees of freedom. The reheating occurs when 
the inflaton begins oscillations. The largest decay rate happens for the largest amplitude 
of oscillations, for instance when $\langle \phi\rangle\sim \phi_0$,~see for 
instance~\cite{AM}. As a result the Universe gets reheated up to a desired temperature:
\begin{equation}
T_R \sim 0.1\sqrt{\Gamma_\phi M_{\rm Pl}}=1.2 \times 10^2 {\rm GeV} \left(\frac{g}{10^{-17}}\right)
\left( \frac{m_\phi}{10^6 {\rm GeV}} \right)^{1/2}\,,
\end{equation}
above the electro-weak scale to facilitate a successful baryogenesis, where 
$\Gamma_\phi=(g^2/8\pi)\left(\langle \phi \rangle /m_\phi \right)^2 m_\phi$ is the
decay rate of $\phi$ with $g$ is the quartic coupling: $g\phi^2 (\eta^\dagger \eta 
+ \chi^\dagger \chi)$.  A small decay rate of inflaton to $\eta^\pm$ and $\chi^\pm$ ensures
the optimal temperature just right above the electro-weak phase transition.

To summarize, we have explored a simple model of lepto-philic universe where DM carries 
a net $B-L$ asymmetry and decays {\it only} into the SM leptons can explain the observed 
positron excess at PAMELA and $e^\pm$ excesses at Fermi. These anomalous observations at 
PAMELA and Fermi may indirectly probe the common origin of mysterious DM and baryon asymmetry as we 
have shown. The baryon asymmetry in our model is created via lepton conserving leptogenesis 
mechanism which gets converted into baryon asymmetry via the electro-weak sphalerons. 
Before closing we note that the model explained here can be embedded within supersymmetry without 
further challenges, neither the leptogenesis nor the DM mechanisms will alter, the parameters and the 
observed values of the inflationary perturbations and the tilt in the spectrum would remain so. 
The details will be published elsewhere~\cite{kmsp}.

Acknowledgement: The authors are supported by the European Union through the Marie
Curie Research and Training Network ``UniverseNet" (MRTN-CT-2006-035863) and  STFC grant, 
PP/D000394/1. NS would like to thank Utpal Sarkar for useful discussions. 



\begin{thebibliography}{99}

\bibitem{PAMELA}
  O.~Adriani {\it et al.},
  arXiv:0810.4995 [astro-ph];

\bibitem{cosmicray_expts}
J.~Chang {\it et al.}, Nature {\bf 456}, 362 (2008);
S.~Torii {\it et al.},
  arXiv:0809.0760 [astro-ph];
F.~Aharonian {\it et.al.},~[HESS Collaboration],
  arXiv:0905.0105 [astro-ph.HE];
J.~J.~Beatty {\it et al.},
  Phys.\ Rev.\ Lett.\  {\bf 93} (2004) 241102;
M.~Aguilar {\it et al.}  [AMS-01 Collaboration],
  Phys.\ Lett.\  B {\bf 646}, 145 (2007).

\bibitem{Fermi}
A.~A.~Abdo {\it et.al.},~[Fermi LAT Collaboration],
  arXiv:0905.0025 [astro-ph.HE];

  \bibitem{BMS_09} C.~Balazs, N.~Sahu and A.~Mazumdar,
  JCAP {\bf 0907}, 039 (2009)
  [arXiv:0905.4302 [hep-ph]].

  \bibitem{uncertainty} I.~V.~Moskalenko and A.~W.~Strong,
  Astrophys.\ J.\  {\bf 493}, 694 (1998);
  T.~Delahaye, R.~Lineros, F.~Donato, N.~Fornengo and P.~Salati,
  Phys.\ Rev.\  D {\bf 77}, 063527 (2008);
E.~A.~Baltz and J.~Edsjo,
  Phys.\ Rev.\  D {\bf 59}, 023511 (1999).
  
  \bibitem{annihilation} 
  L.~Bergstrom, J.~Edsjo and G.~Zaharijas,
  arXiv:0905.0333 [astro-ph.HE];
   M.~Cirelli, M.~Kadastik, M.~Raidal and A.~Strumia,
  arXiv:0809.2409 [hep-ph]\,;
  P.~Meade, M.~Papucci, A.~Strumia and T.~Volansky,
  arXiv:0905.0480 [hep-ph];
  K.~Kohri, J.~McDonald and N.~Sahu,
  arXiv:0905.1312 [hep-ph];
D.~Hooper and T.~M.~P.~Tait,
  arXiv:0906.0362 [hep-ph].
P.~H.~Gu, H.~J.~He, U.~Sarkar and X.~Zhang,
  arXiv:0906.0442 [hep-ph].

  \bibitem{Hamaguchi:2009jb}
  K.~Hamaguchi, K.~Nakaji and E.~Nakamura,
  arXiv:0905.1574 [hep-ph].

  \bibitem{decay} 
  K.~Ishiwata, S.~Matsumoto and T.~Moroi,
  JHEP {\bf 0905}, 110 (2009)
  [arXiv:0903.0242 [hep-ph]].
  A.~Ibarra and D.~Tran,
  JCAP {\bf 0902}, 021 (2009);
   S.~Shirai, F.~Takahashi and T.~T.~Yanagida,
  arXiv:0905.3235 [hep-ph];
  A.~Arvanitaki, S.~Dimopoulos, S.~Dubovsky, P.~W.~Graham, R.~Harnik and S.~Rajendran,
  arXiv:0904.2789 [hep-ph];
  C.~H.~Chen, C.~Q.~Geng and D.~V.~Zhuridov,
  arXiv:0905.0652 [hep-ph];
  N.~Okada and T.~Yamada,
  arXiv:0905.2801 [hep-ph].

\bibitem{astro-orig}
 D.~Hooper,
   P.~Blasi, and
   P.~Dario Serpico,
  Journal of Cosmology and Astro-Particle Physics
  {\bf 1},  25 (2009);
%
 H.~Yuksel,
   M.~D.~Kistler,
  and  T.~Stanev,
  arXiv:0810.2784;
%
 S.~Profumo,
  arXiv:0812.4457;
%
 K.~Ioka,
  arXiv:0812.4851;
%
  E.~Borriello, A.~Cuoco and G.~Miele,
  arXiv:0903.1852 [astro-ph.GA];
%
 P.~Blasi, arXiv:0903.2794; P.~Blasi and P.~D.~Serpico, arXiv:0904.0871
  N.~Kawanaka, K.~Ioka and M.~M.~Nojiri,
  arXiv:0903.3782 [astro-ph.HE].
%
  Y.~Fujita, K.~Kohri, R.~Yamazaki and K.~Ioka,
  arXiv:0903.5298 [astro-ph.HE].


  \bibitem{AEGM}
  R.~Allahverdi, K.~Enqvist, J.~Garcia-Bellido and A.~Mazumdar,
  Phys.\ Rev.\ Lett.\  {\bf 97}, 191304 (2006)
  R.~Allahverdi, K.~Enqvist, J.~Garcia-Bellido, A.~Jokinen and A.~Mazumdar,
  JCAP {\bf 0706}, 019 (2007)

  \bibitem{AKM}
  R.~Allahverdi, A.~Kusenko and A.~Mazumdar,
  JCAP {\bf 0707}, 018 (2007)

\bibitem{ADM}
R.~Allahverdi, B.~Dutta and A.~Mazumdar,
  Phys.\ Rev.\  D {\bf 75}, 075018 (2007)
R.~Allahverdi, B.~Dutta and A.~Mazumdar,
  Phys.\ Rev.\ Lett.\  {\bf 99}, 261301 (2007)


\bibitem{Linde}
A.D. Linde, Particle Physics and Inßationary Cosmol-
ogy (Harwood Academic Publishers, Chur, Switzerland
1990).

\bibitem{yanagida} M.~Fukugita and T.~Yanagida, Phys. Lett. B 174 45,1986.

\bibitem{KMN}
  K.~Kohri, A.~Mazumdar and N.~Sahu,
  arXiv:0905.1625 [hep-ph].


\bibitem{darkmatter&leptogenesis_1} M.~Aoki, S.~Kanemura and O.~Seto,
  Phys.\ Rev.\ Lett.\  {\bf 102}, 051805 (2009);
K.~S.~Babu and E.~Ma,
  Int.\ J.\ Mod.\ Phys.\  A {\bf 23}, 1813 (2008);
T.~Hambye, K.~Kannike, E.~Ma and M.~Raidal,
  Phys.\ Rev.\  D {\bf 75}, 095003 (2007);
E.~Ma,
  Mod.\ Phys.\ Lett.\  A {\bf 21}, 1777 (2006);
  P.~H.~Gu and U.~Sarkar,
  Phys.\ Rev.\  D {\bf 77}, 105031 (2008)\,;
N.~Sahu and U.~A.~Yajnik,
  Phys.\ Lett.\  B {\bf 635}, 11 (2006)
  [arXiv:hep-ph/0509285]\,.


\bibitem{darkmatter&leptogenesis_2} N.~Sahu and U.~Sarkar,
  Phys.\ Rev.\  D {\bf 76}, 045014 (2007);
  J.~McDonald, N.~Sahu and U.~Sarkar,
  JCAP {\bf 0804}, 037 (2008);
N.~Sahu and U.~Sarkar,
  Phys.\ Rev.\  D {\bf 78}, 115013 (2008).


\bibitem{hambye&ma}M.~Frigerio, T.~Hambye and E.~Ma,
  JCAP {\bf 0609}, 009 (2006)
  [arXiv:hep-ph/0603123].

\bibitem{dirac_lep} In case of ``Dirac Leptogenesis" there is no need of Lepton number 
violation. See for instance K. Dick, M. Lindner, M. Ratz and D.Wright, Phys. Rev. Lett.84, 
4039 (2000). 

\bibitem{ma&sarkar}E.~Ma and U.~Sarkar,
  Phys.\ Rev.\ Lett.\  {\bf 80}, 5716 (1998)
  [arXiv:hep-ph/9802445].

\bibitem{LYTH}
J.~C.~Bueno Sanchez, K.~Dimopoulos and D.~H.~Lyth,
  JCAP {\bf 0701} (2007) 015;
 R.~Allahverdi, B.~Dutta and A.~Mazumdar,
  Phys.\ Rev.\  D {\bf 78}, 063507 (2008);
  R.~Allahverdi and A.~Mazumdar,
  arXiv:hep-ph/0610069.

\bibitem{Burgess}
C.~P.~Burgess, R.~Easther, A.~Mazumdar, D.~F.~Mota and T.~Multamaki,
  JHEP {\bf 0505}, 067 (2005).

  \bibitem{AM}
  R.~Allahverdi and A.~Mazumdar,
  JCAP {\bf 0610}, 008 (2006).

\bibitem{Sjostrand:2006za}
  T.~Sjostrand, S.~Mrenna and P.~Skands,
  JHEP {\bf 0605}, 026 (2006).


\bibitem{GCnu}
  J.~Hisano, M.~Kawasaki, K.~Kohri and K.~Nakayama,
  Phys. Rev. D79 (2009) 043516;
  J.~Hisano, K.~Nakayama and M.~J.~S.~Yang,
  arXiv:0905.2075 [hep-ph]; 
J.~Liu, P.~f.~Yin and S.~h.~Zhu,
  arXiv:0812.0964 [astro-ph].


\bibitem{Cowen:2008zz}
  D.~F.~Cowen  [IceCube Collaboration],
  J.\ Phys.\ Conf.\ Ser.\  {\bf 110}, 062005 (2008).

\bibitem{Kappes:2007ci}
  A.~Kappes and f.~t.~K.~Consortium,
  arXiv:0711.0563 [astro-ph].

\bibitem{kmsp} K.~Kohri, A.~Mazumdar,N.~Sahu and P.~Stephens, under preparation.


\end{thebibliography}
\end{document}